\documentclass[12pt]{article}
\usepackage{epsfig,amsfonts,amsthm}
\usepackage{amsmath,amssymb}
\newcommand{\be}{\begin{equation}}
\newcommand{\ee}{\end{equation}}
\newcommand{\bea}{\begin{eqnarray}}
\newcommand{\eea}{\end{eqnarray}}

\newcommand{\doublet}[2]{ \left(\!\! \begin{array}{c}#1 \\ #2 \end{array}\!\!\right) }

\newcommand{\lr}[1]{ \langle #1 \rangle}

\newcommand{\Z}{\mathbb{Z}}

\def\lsim{\mathrel{\rlap{\lower4pt\hbox{\hskip1pt$\sim$}}
    \raise1pt\hbox{$<$}}}         
\def\gsim{\mathrel{\rlap{\lower4pt\hbox{\hskip1pt$\sim$}}
    \raise1pt\hbox{$>$}}}         

\title{$Z_p$ scalar dark matter from multi-Higgs-doublet models}

\author{I.~P.~Ivanov$^{1,2}$, V.~Keus$^1$
\\
  {\small $^1$ IFPA, Universit\'{e} de Li\`{e}ge, All\'{e}e du 6 Ao\^{u}t 17, b\^{a}timent B5a, 4000 Li\`{e}ge, Belgium}\\
  {\small $^2$ Sobolev Institute of Mathematics, Koptyug avenue 4, 630090, Novosibirsk, Russia}\\
  }

\begin{document}
\maketitle

\begin{abstract}
In many models, stability of dark matter particles is protected by a conserved $\Z_2$ quantum number.
However dark matter can be stabilized by other discrete symmetry groups,
and examples of such models with custom-tailored field content have been proposed.
Here we show that electroweak symmetry breaking models with $N$ Higgs doublets can readily accommodate
scalar dark matter candidates stabilized by groups $\Z_p$ with any $p \le 2^{N-1}$, 
leading to a variety of kinds of microscopic dynamics in the dark sector.
We give examples in which semi-annihilation or multiple semi-annihilation processes are allowed or forbidden,
which can be especially interesting in the case of asymmetric dark matter.
\end{abstract}

\section{Introduction}

Despite compelling astronomical evidence for existence of dark matter \cite{DM}, there is still no direct
experimental clue of which particle can play the role of dark matter candidate.
It is only known that dark matter cannot be satisfactorily explained by the Standard Model (SM) particle content.
In this situation, one can focus on exploring dark matter candidates arising in various models beyond the SM,
especially if these models can simultaneously address other particle physics issues such as electroweak symmetry breaking 
and small neutrino masses.

Dark matter particles must be (almost) stable on cosmological timescales.
In many models this stability is provided by a conserved $\Z_2$
quantum number generically called parity. In supersymmetric models its role is played by the $R$-parity,
while in more phenomenologically oriented models such as the Inert doublet model, \cite{inert},
or the minimal singlet model, \cite{singlet},
the $\Z_2$ symmetry is imposed by hand when constructing the lagrangian.
In these models, all the SM particles including the SM-like Higgs boson
have positive parity, while the dark sector particles are of negative parity. 
The lightest among these negative parity particles, which we will generically denote as $d$, 
is stable and represents the dark matter candidate.

It is natural to ask whether dark matter can be stabilized by a conserving discrete quantum number taking values
in a group other than $\Z_2$. This idea was explored in a descent number of works, in which
both abelian \cite{MaZ3,ZNdetailed,Z2Z2} and non-abelian \cite{nonabelian} finite groups were used\footnote{If one takes seriously 
the argument that quantum-gravitational effects violate any global discrete symmetry,
one should require that this quantum number arises as a remnant of a $U(1)$ gauge symmetry spontaneously broken
at a high energy scale, \cite{wilczek}.}. Requiring that $d$ transforms non-trivially under the discrete group
implies that it cannot be a truly neutral particle. It also prohibits direct two-particle annihilation $dd \to X_{SM}$, 
where $X_{SM}$ is any set of SM particles.
However more complicated processes involving several $d$'s can take place.
This certainly changes the kinetics of dark matter evolution in the early Universe 
and its relic abundance after the freeze-out. If in addition one assumes that an asymmetry
between $d$ and its antiparticle $d^*$ is generated at a high energy scale
in a way similar to the baryon asymmetry of the usual matter \cite{asymmetric}, 
then the present day behavior of this asymmetric dark matter is not dominated anymore
by $dd^*$ annihilation and can lead to characteristic observational signatures.

One particular class of groups used to stabilize dark matter are cyclic groups $\Z_p$, see \cite{MaZ3,ZNdetailed} and references therein.
With this choice, all fields are characterized by a conserved quantum number which we will call the $\Z_p$ charge $q$
and which is additive modulo $p$. The usual assignment is that all the SM fields including the SM-like Higgs boson have $q = 0$, 
while the dark matter candidates have nonzero $\Z_p$ charges.
This opens up the possibility of novel two-particle processes such as $dd \to d^* X_{SM}$, which was called semi-annihilation 
in \cite{semi-annihilation}, or even multiparticle versions of it, ``multiple semi-annihilation''.
Besides, if the model allows for existence of several dark matter candidates $d_i$ with different charges $q_i$,
then inelastic two-particle processes $d_i d_j \to d_k X_{SM}$ are also possible. 
Such processes, too, have impact on the kinetics of the dark mater abundances in the early Universe, 
\cite{semi-annihilation-evolution}.

Examples of $\Z_p$-stabilized dark sectors often involve a variety of new fields 
which interact via lagrangians designed specifically to incorporate a given symmetry group, see e.g. \cite{ZNdetailed}.
Even when linked to electroweak physics, these models involve new fields with different electroweak quantum numbers 
(doublets, singlets, etc.), \cite{semi-annihilation-evolution}.
Indeed, if one assumes that extra fields come from a hidden sector coupled to the SM fields via ``portal'' operators \cite{portal}, 
then they must be electroweak singlets by construction.

In this paper we demonstrate that $\Z_p$-stabilized scalar dark matter can easily arise in multi-Higgs-doublet models.
This, perhaps, is not surprising on its own. A less trivial fact is that even with few doublets one can get $\Z_p$
with a rather large $p$.
To be precise, in models with $N$ Higgs doublets, $\Z_p$ with any $p \le 2^{N-1}$ is realizable in the scalar sector.
We will show that this fact can be instrumental in avoiding semi-annihilation processes even with multi-component dark matter sectors.

The structure of the paper is the following. In Section~\ref{section2} we quickly review what is known about symmetries
in the scalar sector of $N$-Higgs-doublet models. Then, in Section~\ref{section3} we give a 3HDM example with 
dark matter candidates stabilized by the $\Z_3$ symmetry group.
In Section~\ref{section4} we show what is possible with four doublets and consider in some detail
the $\Z_7$-symmetric 4HDM, in which one can avoid semi-annihilation processes. 
We end the paper with a discussion and conclusions. 

\section{Scalar sector of NHDM and its symmetries}\label{section2}

The $N$-Higgs-doublet model is a conceptually simple extension of the SM Higgs mechanism.
It is driven by the idea that the Higgs fields, similarly to fermions, can come in several generations.
Its simplest version with only two doublets, 2HDM, is of special interest because it mimics the Higgs sector of MSSM,
and it has been thoroughly studied in the last four decades, see \cite{review2011} and references
therein. In addition, many particular models employing more than two scalar doublets have also been proposed \cite{NHDMvariants}. 

In NHDM one introduces $N$ doublets of complex scalar fields $\phi_i$, $i = 1,\dots, N$, with electroweak isospin $Y=1/2$,
and constructs the self-interaction renormalizable Higgs potential
\be
\label{V:tensorial}
V = Y_{ij}(\phi^\dagger_i \phi_j) + Z_{ijkl}(\phi^\dagger_i \phi_j)(\phi^\dagger_k \phi_l)\,,
\ee
where all indices run from 1 to $N$.
The free parameters of the potential are written as components of tensors $Y_{ij}$ and $Z_{ijkl}$;
in the most general case there are $N^2(N^2+3)/2$ independent parameters.
Once these coefficients and the Yukawa couplings are provided, the model is completely defined 
and the entire phenomenology should follow.
In practice, however, inferring these consequences directly from the lagrangian is impeded by algebraic obstacles at the very first step,
namely, the minimization of a sufficiently generic potential.
The consequence is that only very few general results
are known for $N>2$, \cite{NHDMgeneral-others,abelian3HDM,NHDM2010}.

One particular issue which is of much importance and where certain progress has been recently made 
concerns accidental symmetries which can be encoded
in the scalar sector of NHDM.
These are transformations that mix several Higgs doublets (also called Higgs-basis transformations) but still leave
the potential invariant due to specific patterns in the tensors $Y_{ij}$ and $Z_{ijkl}$.
Although particular models with several Higgs doublets based on various symmetry groups have been proposed and studied
over the last decades, no attempt at systematic classification of possible symmetries was made until very recently.

One of the problems here comes from the observation that just imposing a certain symmetry group $G$ on the Higgs potential
can often lead to potentials symmetric under a {\em larger} group, which includes $G$ as a subgroup.
This feature was discussed in \cite{abelian3HDM} where one-parametric symmetry groups were studied
for 3HDM and in \cite{NHDM2010} where an attempt to understand symmetries in NHDM via geometric constructions in the space
of bilinears was made.
Therefore when describing symmetries of the model, one should focus only on the true symmetry groups, the ones which
are not automatically extended to larger groups. These groups were called in  \cite{NHDM2010} ``realizable groups''.

Unfortunately, no systematic way to reconstruct the true symmetry group of a given NHDM potential for $N>2$ is known so far.
One even does not know the list of realizable symmetry groups possible for $N=3$.
However a step forward was recently made in \cite{abelian2012}, which completely characterized all {\em abelian} groups 
of Higgs-family transformations and generalized $CP$-transformations that can be realized as symmetry groups of the scalar sector of NHDM. 
In particular, in what concerns realizable cyclic groups, it was proved there that for a model with $N$ doublets 
one can construct the Higgs potential symmetric under the group $\Z_p$ with any $1 < p \le 2^{N-1}$
(for $N=3$ this conclusion was known before, \cite{abelian3HDM}).
One can therefore wonder if these models can be used to construct dark matter sectors stabilized by these symmetry groups. 

An electroweak symmetry breaking (EWSB) model with $\Z_p$-stabilized scalar dark matter must satisfy several conditions.
First, the entire lagrangian and not only the Higgs potential must be $\Z_p$-symmetric. 
The simplest way to achieve this is to set the $\Z_p$ charges of all the SM particles to zero and to require
that only one Higgs doublet (the SM-like doublet) couples to fermions. The $\Z_p$ charge of this doublet must be zero,
and it does not matter which doublet is chosen to be SM-like due to the freedom to simultaneously shift the $\Z_p$ charges of all the doublets.
Second, the $\Z_p$ symmetry must remain after EWSB. This is possible when only the SM-like doublet acquires a 
non-zero vacuum expectation value (v.e.v.).
Third, if we insist on $\Z_p$-stabilization, that is, we require that not only decays but also 2-, 3-, $\dots$, $(p-1)$-particle
annihilation to SM fields are forbidden by quantum numbers, then the dark matter candidates
must have $\Z_p$ charge $q$ which is coprime with $p$.

Below we show that multi-Higgs-doublet models can easily satisfy these conditions. 
We start first with the simplest model of this kind, $\Z_3$-symmetric 3HDM, and then 
we outline the plethora of kinds of microscopic dark matter dynamics possible in models with four doublets.

\section{$\Z_3$-symmetric 3HDM}\label{section3}

A natural way to implement the $\Z_3$ symmetry in 3HDM would be to construct a potential symmetric 
under $\phi_1 \to \phi_2 \to \phi_3 \to \phi_1$.
However, upon an appropriate Higgs basis change, this transformation will turn into pure phase rotations of certain doublets.
In fact, it can be proved that {\em any} abelian subgroup of $SU(N)$ can be mapped onto a group of (possibly correlated) 
phase rotations of individual doublets, see \cite{abelian2012} for explicit construction.
Therefore, we will always use below the phase rotation representations of the cyclic symmetry groups.
Also, to keep the notation short, we will describe any such transformation $\phi_i \to e^{i\alpha_i}\phi_i$ by providing 
the $N$-tuple of phases $\alpha_i$.

A scalar potential invariant under a certain group $G$ of phase rotations can be written as a sum $V = V_0 + V_G$,
where $V_0$ is invariant under any phase rotation, while $V_G$ is a collection of extra terms which
realize the chosen symmetry group.
The generic phase rotation invariant part has form
\be
\label{Tsymmetric}
V_0 = \sum_i \left[- m_i^2 (\phi_i^\dagger \phi_i) + \lambda_{ii} (\phi_i^\dagger \phi_i)^2\right] 
+ \sum_{ij}\left[\lambda_{ij}(\phi_i^\dagger \phi_i) (\phi_j^\dagger \phi_j) + 
\lambda'_{ij}(\phi_i^\dagger \phi_j) (\phi_j^\dagger \phi_i)\right]\,,
\ee
while $V_G$ obviously depends on the group. In particular, for the group $\Z_3$ in 3HDM we have
\be
V_{\Z_3} = \lambda_{1}(\phi_3^\dagger\phi_1)(\phi_2^\dagger\phi_1) + 
\lambda_{2}(\phi_1^\dagger\phi_2)(\phi_3^\dagger\phi_2) + 
\lambda_{3}(\phi_2^\dagger\phi_3)(\phi_1^\dagger\phi_3) + h.c.,
\ee
where at least two of the coefficients $\lambda_1,\,\lambda_2,\,\lambda_3$ are non-zero (otherwise, the potential 
would have a continuous symmetry).
This potential is symmetric under the phase rotations generated by
\be
a = {2\pi \over 3}(0,\, 1,\, 2)\,,\quad a^3 = 1\,,\label{generatorZ3}
\ee
In fact, the assignment of these charges to the three doublets is completely arbitrary, and the group generated by
the generator $a$ with permuted charges has the same action of the potential.
In (\ref{generatorZ3}) we simply chose $\phi_1$ to be the SM-like doublet.

Whether this symmetry is conserved or spontaneously broken depends on the pattern of the vacuum expectation values.
If we insist on conservation of the $\Z_3$ symmetry, we require that 
$\lr{\phi_1^0} = v_1 = v/\sqrt{2}$, $\lr{\phi_2^0} = \lr{\phi_3^0} = 0$.
There is nothing suprising that the potential $V_0 + V_{\Z_3}$ can have a $\Z_3$-symmetric global minimum upon an appropriate choice
of coefficients. The question is whether it requires any fine-tuning or not. Below we show that it does not, and the minimum
of the type $(v_1,\,0,\,0)$ arises in a sizable part of the entire available parameter space of this model.

First, we note that if $(v_1,\,0,\,0)$ is an extremum of $V_0$, then it is also an extremum of $V_0 + V_{\Z_3}$ 
because the extra terms contain $\phi_1$ only linearly and quadratically.
Therefore, when constructing a model example, one can first build $V_0$ with a global minimum at $(v_1,\,0,\,0)$
and then add a sufficiently weak $V_{\Z_3}$ so that this point remains a minimum.

Now, turning to minimization of $V_0$, suppose that we search for the neutral minimum with a generic complex 
v.e.v. pattern $(v_1,\, v_2,\, v_3)$.
By introducing $\rho_i = |v_i|^2 \ge 0$, we can rewrite $V_0$ as
\be
V_0 = - M_i \rho_i + {1 \over 2}\Lambda_{ij}\rho_i \rho_j = {1 \over 2}\Lambda_{ij}(\rho_i -\mu_i)(\rho_j - \mu_j) + const\,.\label{shifted}
\ee
Here $M_i = (m_1^2,\,m_2^2,\,m_3^2)$, $\Lambda_{ij}$ is constructed from $\lambda_{ij}$ and $\lambda_{ij}^\prime$
in an obvious way, and $\mu_i = (\Lambda_{ij})^{-1}M_j$.
Positivity condition on the potential guarantees that $\Lambda_{ij}$ is a positive definite matrix, therefore its inverse exists
(we omit here the degenerate situations when the quartic potential has flat directions).

The allowed values of the $\rho_i$ populate the first octant ($\rho_i \ge 0$) in the three-dimensional euclidean space.
Due to (\ref{shifted}), the search for the global minimum can be reformulated as the search for the point in the 
first octant which lies closest to $\mu_i$ in the euclidean metric defined by $\Lambda_{ij}$.
Clearly, if $\mu_i$ itself lies inside the first octant ($\mu_1,\, \mu_2,\, \mu_3 \ge 0$), then the global minimum is at $\rho_i = \mu_i$.
If $\mu_i$ lies outside the first octant, then the closest point lies either on a face, or on an edge of the first octant, or at the origin.
The global minimum is unique by the convexity arguments.
It is now clear from this geometric construction that the entire space of all possible vectors $\mu_i$ can be broken into several regions
of non-zero measure which correspond to all of these possibilities. 
In particular, the region corresponding to vacuum
alignment of the type $(v_1,\,0,\,0)$ also has a non-zero measure and fills a sizable part of the parameter space. 
In this sense, this vacuum pattern does not require any fine-tuning of the coefficients of the potential.

Having established that the required vacuum pattern is generically possible, we now switch to a simple
version of the model. This is done only to simplify the presentation of the argument;
if needed, the calculations can be repeated for a generic potential.
Namely, we now choose $m_1^2 > 0$, $m_2^2,\,m_3^2 < 0$ and take $\Lambda_{ij} = 2\lambda_0 \delta_{ij}$,
so that the potential becomes
\bea
V &=& - m_1^2 (\phi_1^\dagger \phi_1) + |m_2^2| (\phi_2^\dagger \phi_2) + |m_3^2| (\phi_3^\dagger \phi_3)
+ \lambda_0 \left[(\phi_1^\dagger \phi_1)^2+ (\phi_2^\dagger \phi_2)^2 + (\phi_3^\dagger \phi_3)^2\right]\nonumber\\
&& + \lambda_{1}(\phi_3^\dagger\phi_1)(\phi_2^\dagger\phi_1) + 
\lambda_{2}(\phi_1^\dagger\phi_2)(\phi_3^\dagger\phi_2) + 
\lambda_{3}(\phi_2^\dagger\phi_3)(\phi_1^\dagger\phi_3) + h.c.
\eea
By construction, its global minimum is at $\lr{\phi_i^0} = (v/\sqrt{2},\,0,\,0)$, where $v^2 = m_1^2/\lambda_0$.
In order to find the mass matrices, we write the doublets as
\be
\phi_1 = \doublet{G^+}{{1\over\sqrt{2}}(v + h + i G^0)}\,,\quad 
\phi_{2} = \doublet{w_{2}^+}{z_{2}}\,,\quad 
\phi_{3} = \doublet{w_{3}^+}{z_{3}}\,.\label{expand}
\ee
Here $h$ is the SM-like Higgs boson, $G^0$ and $G^+$ are the would-be Goldstone bosons, 
while $w_2^+,\, w_3^+$ and $z_2,\, z_3$ are charged and neutral Higgs bosons, respectively.
Fields $w_{2,3}^+$ and $z_{2,3}$ have well-defined $\Z_3$-charges: $q=1$ for $w_2^+,\, z_2$ as well as for $w_3^-,\, z_3^*$
and $q=2$ ( $=-1$ mod 3) for $w_3^+,\, z_3$ and $w_2^-,\, z_2^*$. 
Note that in contrast to the usual practice, we describe the neutral Higgs bosons in the second and third doublet
by complex fields rather than pairs of neutral fields.
The reason is that, by construction, the fields corresponding to the real and imaginary pairs of $z$'s have identical masses
and coupling constants. In any process that can arise in this model, these two fields are emitted and absorbed simultaneously,
so they can be described by a single complex field $z_i$.

The SM-like Higgs boson has mass $m_h^2 = 2m_1^2$, while the masses of the charged Higgs bosons are 
$m_{w_2^\pm}^2 = |m_2^2|$, $m_{w_3^\pm}^2 = |m_3^2|$.
Neutrals with equal $q$ can mix, which indeed happens at $\lambda_1 \not = 0$. 
We descibe the resulting mass eigenstates by complex fields $d$ and $D$ ($m_d < m_D$), both having $q=1$: 
\bea
&& d = \cos\alpha\, z_2 + \sin\alpha e^{-i\beta}\, z_3^*\,,\quad D = - \sin\alpha e^{i\beta}\, z_2 + \cos\alpha\, z_3^*\,,\nonumber\\
&& \tan2\alpha = {|\lambda_1| \over \lambda_0}{m_1^2 \over |m_2^2|-|m_3^2|}\,,\quad \beta = \mbox{arg}\, \lambda_1\,,\nonumber\\
&& m_{D,d}^2 = {|m_2^2| + |m_3^2| \over 2} \pm {1 \over 2}\sqrt{(|m_2^2| - |m_3^2|)^2 + {|\lambda_1|^2 \over \lambda_0^2}m_1^4}\,.
\eea
Note that within this model we have
\be
m_d < m_{w_2^\pm},\, m_{w_3^\pm} < m_D\,,\quad m_{w_2^\pm}^2 + m_{w_3^\pm}^2 = m_d^2 + m_D^2\,. 
\ee
The triple and quartic interaction terms arising from $V_0$ and $V_{\Z_3}$ 
specify the dynamics of the dark matter candidates.
The lightest particle from the second and third Higgs generations is $d$, and it is stabilized against decaying into the SM particles by the $\Z_3$
symmetry.
As for heavier particles, triple interactions lead to their decays such as $D \to dh$, $D \to dZ$, and $D \to w_2^+ W^-$, $D \to w_3^- W^+$ if allowed kinematically.
If the mass splitting between the $d$ and $D$ is small, then these processes involve virtual $h$, $Z$, etc. which then decay into the SM particles.
In this aspect, $D$ decays are similar to weak decays. Charged Higgs bosons $w_{2,3}^\pm$ will also decay to $d$ or $d^*$ plus SM particles.

In the case of symmetric dark matter, the main process leading to depletion of dark matter after electroweak symmetry breaking 
is the direct annihilation $dd^* \to X_{SM}$, and the semi-annihilation reaction discussed below is only a correction to this process.
Still, it might be possible that this correction leads to a sizable departure of the kinetics of the dark matter in the early Universe
and affects the relic abundance at the freeze-out, \cite{semi-annihilation-evolution}.

The situation becomes more interesting in models of asymmetric dark matter, \cite{asymmetric},
in which an asymmetry between $d$ and $d^*$ is generated at a higher energy scale.
It is possible for example that upon electroweak phase transition almost all $d$'s annihilate with $d^*$ into the SM sector, 
leaving behind a certain concentration of dark matter candidates $d$. 
In the present epoch, $d$ can scatter elastically, $dd \to dd$ with $\sigma_{el.} \propto \lambda_0^2$, 
but they can also initiate semi-annihilation processes such as
$dd \to d^*X_{SM}$ with a subsequent annihilation of $d^*$ with a $d$.
This possibility originates from the following quartic terms in the scalar potential 
\be
{1\over \sqrt{2}} h ddd \cos\alpha \sin\alpha e^{i\beta} (\lambda_2 \cos\alpha + \lambda_3^*\sin\alpha e^{i\beta}) + h.c.
\ee
Depending on $\lambda$'s, this process can be as efficiently as the direct annihilation 
in the usual annihilating dark matter models, or it can be suppressed by the small coupling constant. 

Finally, the same interaction terms also generate the triple annihilation processes $ddd \to h \to X_{SM}$,
whose rate is, however, suppressed at small densities with respect to the semi-annihilation.

\section{Avoiding semi-annihilation}\label{section4}

The presence of the $hddd$ terms in the interaction lagrangian in the previous example, which were responsible for the two-particle
semi-annihilation process, was due to the $\Z_3$ symmetry group. 
One can wonder whether two-particle semi-annihilation can be avoided by employing a $\Z_p$ group with larger $p$.
In this present section we show that it is indeed possible in a model with four Higgs doublets.

According to \cite{abelian2012}, one can encode in the 4HDM scalar sector any group $\Z_p$ with $p \le 8$.
Note that in order to avoid continuous symmetry, one must accompany the phase-symmetric part of the potential $V_0$
with at least three distinct terms transforming non-trivially under phase rotations.
In Table 1 we give the list of these symmetry groups together with examples of the three interaction terms and the phase rotations that
generate the corresponding group.

\begin{table}[!htb]
\begin{center}
\begin{tabular}{ c c c }	
\hline
group & interaction terms & phase rotations \\
\hline
  $\Z_2$ &  $(1^\dagger 2),\  (1^\dagger 3),\ (1^\dagger 4)^2$ & ${2\pi \over 2}(0,\,0,\,0,\,1)$  \\[2mm]
  $\Z_3$ &  $(3^\dagger 2),\  (1^\dagger 3)(4^\dagger 3),\ (1^\dagger 4)(1^\dagger 2)$ & ${2\pi \over 3}(0,\,1,\,1,\,2)$  \\[2mm]
  $\Z_4$ &  $(3^\dagger 2),\  (1^\dagger 3)(4^\dagger 3),\ (1^\dagger 4)^2$ & ${2\pi \over 4}(0,\,1,\,1,\,2)$  \\[2mm]
  $\Z_5$ &  $(4^\dagger 3)(2^\dagger 3),\  (3^\dagger 2)(1^\dagger 2),\ (4^\dagger 1)(3^\dagger 1)$ & ${2\pi \over 5}(0,\,1,\,2,\,3)$  \\[2mm]
  $\Z_6$ &  $(4^\dagger 3)(2^\dagger 3),\  (3^\dagger 2)(1^\dagger 2),\ (1^\dagger 4)^2$ & ${2\pi \over 6}(0,\,1,\,2,\,3)$  \\[2mm]
  $\Z_7$ &  $(4^\dagger 1)(3^\dagger 1),\  (4^\dagger 3)(2^\dagger 3),\ (4^\dagger 2)(1^\dagger 2)$ & ${2\pi \over 7}(0,\,2,\,3,\,4)$  \\[2mm]
  $\Z_8$ &  $(4^\dagger 3)(2^\dagger 3),\  (4^\dagger 2)(1^\dagger 2),\ (1^\dagger 4)^2$ & ${2\pi \over 8}(0,\,2,\,3,\,4)$  \\
\hline
\end{tabular}
\caption{Cyclic groups realizable as symmetry groups in the scalar sector of 4HDM; $(a^\dagger b)$ is a short notation for $(\phi^\dagger_a\phi_b)$.}
\end{center}
\end{table}

Two remarks concerning this table are in order.
First, we stress that all these groups are realizable, so that if the three terms in each line are written down with non-zero coefficients,
there is no phase transformation other than multiple of the generator and the common overall phase shift that leaves them invariant.
Second, since the potential is invariant under the common phase shift of all doublets, one can
freely add additional equal phases to the generators shown in the third column and, in addition, one can permute the doublets. 
For example, the last line of this table can be replaced by $(3^\dagger 2)(1^\dagger 2),\  (3^\dagger 1)(4^\dagger 1),\ (4^\dagger 3)^2$,
which is symmetric under the $\Z_8$ group generated by phase rotations ${2\pi \over 8}\, (0,\, 1,\, 2,\, 6)$. 

The patterns of phase shifts given in this table allow for construction of various dark sectors with different possibilities
for dark matter dynamics. Here, we do not aim at a complete classification of these possibilities but would like only to show
that there are examples in 4HDM in which semi-annihilation of dark matter candidates is also forbidden.

To this end, let us consider the $\Z_7$-symmetric 4HDM with the potential $V = V_0 + V_{\Z_7}$ with 
\be
V_{\Z_7} = \lambda_{1}(\phi_4^\dagger\phi_1)(\phi_3^\dagger\phi_1) + 
\lambda_{2}(\phi_4^\dagger\phi_2)(\phi_1^\dagger\phi_2) + 
\lambda_{3}(\phi_4^\dagger\phi_3)(\phi_2^\dagger\phi_3) + h.c. \label{VZ7}
\ee
As before, we assume that only the first doublet couples to fermions.
Using the technique of the previous section,
one can easily construct the potential $V_0$ with the global minimum at $(v/\sqrt{2},\,0,\,0,\,0)$.
Then expanding the doublets similarly to (\ref{expand}), one observes that $z_3$ and $z_4^*$
have $\Z_7$-charges $q = 3$ and mix via the $\lambda_1$ term leading to mass eigenstates $d$ and $D$.
In addition, we have the field $z_2$ with $\Z_7$-charge $q= 2$ and electrically charged Higgs bosons $w_{2,3,4}^\pm$
with appropriate $\Z_7$-charges.

By adjusting free parameters, one can easily make $d$ the lightest among the particles that transform
non-trivially under $\Z_7$. Then, the other particles will either eventually decay to $d$ or $d^*$ plus SM particles or 
will be stable representing an additional contribution to dark matter.
Again, if the asymmetry between $d$ or $d^*$ exists and if the rate of their annihilation is high, 
then after the freeze-out we are left with the gas predominantly made of $d$'s.

The subsequent microscopic dynamics depends on the interactions between $d$'s and $z_2$'s which follow from (\ref{VZ7}).
The relevant terms are $d z_2 z_2$ times a SM field from the $\lambda_2$ term and $dddz_2^*$ from the $\lambda_3$ term.
One- or two-particle processes such as $d\to z_2^* z_2^*$, $dd \to d^*z_2$, and $dd \to z_2z_2z_2$ are 
all kinematically forbidden.
Multiple collision kinetics depends on whether $m_{z_2} < 3 m_d$ or not.
If $z_2$ is not too heavy, then the ``triple semi-annihilation'', $ddd \to z_2 X_{SM}$, is kinematically allowed
and will create a population of $z_2$ even if it were absent before. 
However, $z_2$ will get depleted by the semi-annihilation process $z_2 z_2 \to d^* X_{SM}$.
So, if one starts with a certain concentration of $z_2$, $d$ and their antiparticles, then $z_2$ will die off at a higher rate than $d$'s.
In stationary conditions, the terminal concentrations will be those equilibrating the rates of the following $6d$ tree-level scattering
with intermediate $z_2$'s:
\be
6d \to z_2 z_2 X_{SM}\to d^* X'_{SM}\,,\label{6d}
\ee
and the subsequent annihilation of $d^*$. The net result of this chain will be the ``$7d$-burning process'', $7d \to X_{SM}$,
the bottleneck in this chain being the triple-$d$ process $ddd \to z_2 X_{SM}$.

On the other hand, if $m_{z_2} > 3 m_d$, then $ddd \to z_2$ is kinematically forbidden, while the inverse process leads to a quick $z_2$ decay. 
In this case, one can still burn $d$'s via the tree-level process with intermediate virtual $z_2$'s: 
\be
dddd \to d^*d^* z_2 z_2 \to d^*d^*d^* X_{SM}\,,\label{4d}
\ee
The net result will be the same $7d$-burning, but the bottleneck process is now the $4d$ collision, whose rate is even stronger suppressed.

\section{Discussion and conclusions}\label{section5}

The main purpose of this paper is to demonstrate that there already exists a phenomenological template for 
scalar dark matter models stabilized by cyclic groups larger than $\Z_2$.
This template uses several electroweak Higgs doublets decoupled from fermions, and
it represents one of the simplest extensions of the Standard Model.
Remarkably, models with few doublets can easily accommodate dark sectors which are stabilized by a large list of discrete groups
and which display various kinds of microscopic dynamics. In particular, we gave explicit examples of dark sectors where the bottleneck process 
leading to depletion of asymmetric dark matter can be a 2-particle, 3-particle or 4-particle semi-annihilation.
We stress that these models do not require any serious fine-tuning. We only ask for the presence of terms invariant under
the chosen symmetry group but do not constrain coefficients in front of these terms.

In certain aspects these models resemble the Inert Doublet Model, \cite{inert}, but in the other they
rely on symmetry patterns that arise only with several doublets.
In this respect, such models can be viewed as ``multi-inert'' doublet models although this name of course does not completely
specify the microscopic dynamics. 

Exploring the observational consequences of each sort of microscopic dynamics is a separate task.
It should include study of the dark matter kinetics in two situations. First, one obviously needs to track down the dark matter evolution 
in the expanding Unverse after the electroweak phase transition and determine the freeze-out abundances.
Analysis of \cite{semi-annihilation-evolution} already proves that semi-annihilation processes can be important,
but it remains to be understood how sensitive the evolution is to the exact microsopic dynamics.

We would like to stress that studying this problem in the context of multi-Higgs-doublet models can be much subtler than it looks at first glance
due to multiple phase transitions near and below the electroweak scale.
Indeed, even in the two-Higgs-doublet model a single electroweak phase transition can split into a sequence of several phase transitions
of different nature, both in the general case \cite{thermal} and in the Inert doublet model \cite{thermalIDM}. 
One can expect that even longer chains of phase transitions can be possible in multi-doublet models.
Note that the last among these phase transition can in principle happen at temperatures much lower than the nominal electroweak
temperature scale. Consequently, the Universe might have evolved through a sequence of vacua with different, and perhaps exotic, properties.
Phase transitions between these phases could have led to complete restructuring of the particle mass spectrum, both within the SM and in the dark sector;
particles which are stable in one phase can be unstable in another. All these delicate details, as well as the thermodynamics of the phase transitions
themselves, can modify the evolution of the dark sector. None of the existing evolution codes can adequately address these intricacies.

The second situation where the microscopic dark matter dynamics can make an impact is the present epoch
evolution at astrophysical sites of elevated dark matter concentrations (galactic centers, interiors of compact stars, etc., \cite{accumulation}). 
Since multi-particle processes are involved, the sensitivity to the dark matter density will be different
from that of the usual two-particle annihilating or with truly non-annihilating dark matter.

For example, it is known that dark matter with sufficient elastic cross section can get captured inside
neutron stars, \cite{accumulation}. In models with asymmetric truly non-annihilating scalar dark matter,
its accumulation can lead to formation of the Bose-Einstein condensate 
(which means that the occupation number in the phase space can become large) 
or even to collapse in a tiny black hole \cite{constraints}.
In a certain region of parameter space, this black hole will destroy the host neutron star sufficiently
quickly compared to the typical neutron star lifetime; therefore, this region is excluded by observations.
In the case of multi-inert dark matter accumulated inside a neutron star,
multiple annihilation processes will effectively enter the game as the density reaches a certain threshold,
precluding black hole formation and avoiding the above constraints.\\

In conclusion, we showed that multi-Higgs-doublet models can naturally accommodate
scalar dark matter candidates protected by the group $\Z_p$. For a model with $N$ doublets,
the values of $p$ can be as large as $2^{N-1}$.
These models do not require any significant fine-tuning and can lead to a variety of forms of microscopic dynamics
among the dark matter candidates (allowing or forbidding semi-annihilation,
offering different routes to multi-particle annihilation, etc.).

\section*{Acknowledgements} 
This work was support by the Belgian Fund F.R.S.-FNRS via the
contract of Charg\'e de recherches, and in part by grants
RFBR No.11-02-00242-a and NSh-3810.2010.2.


\begin{thebibliography}{99}
\bibitem{DM} 
  G.~Bertone, D.~Hooper and J.~Silk,
  Phys.\ Rept.\  {\bf 405}, 279 (2005).
\bibitem{inert}
N.~G.~Deshpande and E.~Ma,
  Phys.\ Rev.\ D {\bf 18}, 2574 (1978);
L.~Lopez Honorez, E.~Nezri, J.~F.~Oliver and M.~H.~G.~Tytgat,
  JCAP {\bf 0702}, 028 (2007).
\bibitem{singlet}
  C.~P.~Burgess, M.~Pospelov and T.~ter Veldhuis,
  Nucl.\ Phys.\ B {\bf 619}, 709 (2001).
\bibitem{MaZ3}
E.~Ma,
  Phys.\ Lett.\ B {\bf 662}, 49 (2008).
\bibitem{ZNdetailed}
B.~Batell,
  Phys.\ Rev.\ D {\bf 83}, 035006 (2011).
\bibitem{Z2Z2}
G.~Belanger and J.~-C.~Park,
  arXiv:1112.4491 [hep-ph].
\bibitem{nonabelian}
A.~Adulpravitchai, B.~Batell and J.~Pradler,
  Phys.\ Lett.\ B {\bf 700}, 207 (2011).
\bibitem{wilczek}
L.~M.~Krauss and F.~Wilczek,
  Phys.\ Rev.\ Lett.\  {\bf 62}, 1221 (1989).
\bibitem{asymmetric}
S.~Nussinov,
  Phys.\ Lett.\ B {\bf 165}, 55 (1985);
D.~B.~Kaplan,
  Phys.\ Rev.\ Lett.\  {\bf 68}, 741 (1992);
D.~E.~Kaplan, M.~A.~Luty and K.~M.~Zurek,
  Phys.\ Rev.\ D {\bf 79}, 115016 (2009).
\bibitem{semi-annihilation} 
  F.~D'Eramo and J.~Thaler,
  JHEP {\bf 1006}, 109 (2010).
\bibitem{semi-annihilation-evolution} 
G.~Belanger, K.~Kannike, A.~Pukhov and M.~Raidal,
  arXiv:1202.2962 [hep-ph].
\bibitem{portal} 
  B.~Patt and F.~Wilczek,
  hep-ph/0605188.
\bibitem{review2011}
G.~C.~Branco et al,   
arXiv:1106.0034 [hep-ph].
\bibitem{NHDMvariants}
S.~Weinberg,  
 Phys.\ Rev.\ Lett.\  {\bf 37}, 657 (1976);
S.~L.~Adler,
  Phys.\ Rev.\ D {\bf 60}, 015002 (1999);
R.~A.~Porto and A.~Zee,
  Phys.\ Lett.\ B {\bf 666}, 491 (2008);
A.~C.~B.~Machado, J.~C.~Montero and V.~Pleitez,
  Phys.\ Lett.\ B {\bf 697}, 318 (2011).
\bibitem{NHDMgeneral-others}
L.~Lavoura and J.~P.~Silva,
  Phys.\ Rev.\  D {\bf 50}, 4619 (1994).
N.~G.~Deshpande and X.~G.~He,  
  Pramana {\bf 45}, S73 (1995);
R.~Erdem,  
  Phys.\ Lett.\  B {\bf 355}, 222 (1995);
A.~Barroso, P.~M.~Ferreira, R.~Santos and J.~P.~Silva,  
  Phys.\ Rev.\  D {\bf 74}, 085016 (2006);
C.~C.~Nishi,  
  Phys.\ Rev.\  D {\bf 76} (2007) 055013;
K.~Olaussen, P.~Osland and M.~A.~.Solberg,
  JHEP {\bf 1107}, 020 (2011).
\bibitem{abelian3HDM} P.~M.~Ferreira and J.~P.~Silva,
  Phys.\ Rev.\  D {\bf 78}, 116007 (2008).
\bibitem{NHDM2010}I.~P.~Ivanov, C.~C.~Nishi, Phys.\ Rev.\ D {\bf 82}, 015014 (2010);
I.~P.~Ivanov, JHEP {\bf 1007}, 020 (2010).
\bibitem{abelian2012}
I.~P.~Ivanov, V.~Keus,  E.~Vdovin, arXiv:1112.1660 [math-ph].
\bibitem{thermal}
I.~P.~Ivanov,
  Acta Phys.\ Polon.\ B {\bf 40}, 2789 (2009);
I.~F.~Ginzburg, I.~P.~Ivanov and K.~A.~Kanishev,
  Phys.\ Rev.\ D {\bf 81}, 085031 (2010).
\bibitem{thermalIDM}
I.~F.~Ginzburg, K.~A.~Kanishev, M.~Krawczyk and D.~Sokolowska,
  Phys.\ Rev.\ D {\bf 82}, 123533 (2010).
\bibitem{accumulation} 
I.~Goldman and S.~Nussinov,
  Phys.\ Rev.\ D {\bf 40}, 3221 (1989);
C.~Kouvaris,
  Phys.\ Rev.\ D {\bf 77}, 023006 (2008);
G.~Bertone and M.~Fairbairn,
  Phys.\ Rev.\ D {\bf 77}, 043515 (2008);
A.~de Lavallaz and M.~Fairbairn,
  Phys.\ Rev.\ D {\bf 81}, 123521 (2010);
T.~Guver, A.~E.~Erkoca, M.~H.~Reno and I.~Sarcevic,
  arXiv:1201.2400 [hep-ph].
\bibitem{constraints}
C.~Kouvaris and P.~Tinyakov,
  Phys.\ Rev.\ D {\bf 82}, 063531 (2010);
 C.~Kouvaris and P.~Tinyakov,
  Phys.\ Rev.\ D {\bf 83}, 083512 (2011);
S.~D.~McDermott, H.~-B.~Yu and K.~M.~Zurek,
  Phys.\ Rev.\ D {\bf 85}, 023519 (2012).

\end{thebibliography}
\end{document}